\documentclass[11pt ]{article}
\usepackage{amssymb}
\usepackage{amsmath}
\usepackage{enumitem}
\usepackage[utf8]{inputenc}
\usepackage{hyperref}
\hypersetup{hidelinks}
\usepackage{color}
\usepackage{cite} 
\usepackage{soul}
\usepackage[symbol]{footmisc}

\makeatletter
\def\@xfootnote[#1]{%
  \protected@xdef{#1}%
  \@footnotemark\@footnotetext}
\makeatother

 \newcommand\snowmass{\begin{center}\rule[-0.2in]{\hsize}{0.01in}\\\rule{\hsize}{0.01in}\\
\vskip 0.1in Submitted to the  Proceedings of the US Community Study\\
on the Future of Particle Physics (Snowmass 2021)\\
\rule{\hsize}{0.01in}\\\rule[+0.2in]{\hsize}{0.01in} \end{center}}

\usepackage{geometry}
 
\begin{document}
 \snowmass
 \centerline{\bf \large  Hadron Spectroscopy with Lattice QCD\footnote{The White Paper of the topical group Hadron Spectroscopy (RF7) within the Snowmass 2021; submitted also to the topical group Lattice Gauge Theory (TF05)}}
  
  \vspace{0.7cm}
  
  \centerline{John Bulava$^{1}$,  Ra\'ul  Brice\~no$^{2,3}$, William Detmold$^{4,5}$, Michael D\"oring$^{6,2}$, Robert G. Edwards$^{2}$, }
  
  \vspace{0.1cm}
  
  \centerline{  Anthony Francis$^{7,8,9}$, Francesco Knechtli$^{10}$, Randy Lewis$^{11}$,  Sasa Prelovsek$^{12,13,}$\footnote{Contact person: Sasa Prelovsek, sasa.prelovsek@ijs.si},  }

\vspace{0.1cm}

 \centerline{  Sin\'ead M. Ryan$^{14}$, Akaki Rusetsky$^{15,16}$, Stephen R.  Sharpe$^{17}$, Adam Szczepaniak$^{2,18,19}$,   }
 
 \vspace{0.1cm}

 \centerline{ Christopher E. Thomas$^{20}$, Michael L. Wagman$^{21}$, Marc Wagner$^{22,23}$}
 
\vspace{0.5cm}
 
 {\small \it
  $^{(1)}$ Deutsches Elektronen-Synchrotron DESY, Zeuthen, Germany
    
  $^{(2)}$ Thomas Jefferson National Accelerator Facility, Virginia, USA 
    
 $^{(3)}$ Department of Physics, Old Dominion University,  Newport News, USA  
   
   $^{(4)}$ Center for Theoretical Physics, Massachusetts Institute of Technology, Cambridge, USA 
   
    $^{(5)}$ Institute for Artificial Intelligence and Fundamental Interactions, USA
  
  $^{(6)}$  George Washington University, Washington,  USA
    
$^{(7)}$   Albert Einstein Center, Universit\" at Bern,   Switzerland

$^{(8)}$ Institute of Physics, National Yang Ming Chiao Tung University,   Taiwan

$^{(9)}$ Theory Department, CERN,     Switzerland

$^{(10)}$  Deptartment of Physics, University of Wuppertal, Germany

$^{(11)}$ Department of Physics and Astronomy, York University, Toronto,  Canada

$^{(12)}$ Faculty of Mathematics and Physics, University of Ljubljana, Slovenia

$^{(13)}$ Jozef Stefan Institute,  Slovenia

$^{(14)}$   School of Mathematics and Hamilton Mathematics Institute,

$\quad\ $ Trinity College, Dublin 2, Ireland

$^{(15)}$ HISKP Theory, Rheinische Friedrich-Wilhelms-Universit\" at Bonn,   Germany 

$^{(16)}$ Tbilisi State University,   Georgia

$^{(17)}$ Physics Department, University of Washington, Seattle,  USA

$^{(18)}$  Physics Department, Indiana University, Bloomington,  USA

 $^{(19)}$ Center for Exploration of Energy and Matter, Indiana University, Bloomington,  USA 
  
 $^{(20)}$   Department of Applied Mathematics and Theoretical Physics, University of Cambridge,  UK
 
  $^{(21)}$ Fermi National Accelerator Laboratory, Batavia, USA
    
 $^{(22)}$   Goethe-Universi\" at Frankfurt am Main,  Germany
  
 $^{(23)}$  Helmholtz Research Academy Hesse for FAIR,  Frankfurt am Main, Germany

   }

  \vspace{1cm}
  
  \begin{minipage}[c]{0.9\hsize}
  {\bf Abstract:} The status  and prospects for investigations  of exotic and conventional hadrons with lattice QCD are discussed.  The majority of hadrons decay strongly via one or multiple decay-channels, including most of the experimentally discovered exotic hadrons. 
Despite this difficult challenge, the properties of several hadronic resonances have been determined within lattice QCD. To further discern the spectroscopic properties of various hadrons and to help resolve their nature we present our suggestions for future analytic and lattice studies.
 \end{minipage}
 
   \vspace{1cm}
 
 
 \tableofcontents
 
 \newpage
 
\section{   Executive Summary }
  
  There has been a renaissance in hadron spectroscopy in the last twenty years sparked by experimental discovery
of strongly-interacting exotic matter and motivated by the desire to understand the nature and structure of
predicted and observed hadronic resonances in terms of fundamental forces and interactions. 

Lattice Quantum Chromodynamics (LQCD) is the formulation of QCD on a discrete space-time grid which
enables first principles, systematically improvable, numerical simulations of strong interaction physics.
 The approach offers, in principle, the possibility to determine strong-interaction physics with
 quantifiable uncertainties thereby complementing and supporting experimental searches with theoretical insight. 
  In a lattice calculation there can be a subtle interplay between the systematic uncertainties that requires
 theoretical understanding and these may be quantity-dependent. In addition, as discussed in this paper,
 the finite volume and the quark mass dependence can be used to advantage in lattice hadronic physics.
 
 In the last decade the precision and predictive power of lattice hadron spectroscopy has dramatically improved. Many
 bound states, stable under strong decay, are determined at sub-percent statistical precision and with all systematic
 uncertainties quantified or removed.
 However, it is in calculations of states close to or above strong decay thresholds where the most significant
 progress has been made. Underpinned by advances in algorithms and analytic understanding, lattice calculations of
 conventional and exotic light- and heavy-quark hadronic resonances are already having a significant impact on our
 understanding of hadron spectroscopy. 

 \begin{itemize}
 \item 
   The processes $\rho\rightarrow\pi\pi$ and $K^\ast\rightarrow K\pi$ are now well-studied by multiple lattice
   groups and serve as benchmarks for methods. It is worth noting that lattice simulations at the physical pion mass
   are possible although costly and can lead to a proliferation of decay channels that complicate the extraction of
   resonance parameters. Nevertheless, we expect that the light quark mass dependence of many processes
   including e.g. scalar resonances, will be
   mapped out in the next five years providing valuable information on resonance structure and calculations
   that are directly relevant for experiments.

\item Most exotic states of interest decay to multiple hadronic final states and/or to states with non-zero spin hadrons. There have been early,
   pioneering lattice calculations in the light sector; however, in the heavy-quark sector only coupled
   $D\pi-D\eta-D_s\bar{K}$ and $D\bar D-D_s\bar D_s$ have been tackled to date. An important goal in the next five years for LQCD is
   the physics of heavy quark resonance scattering in coupled-channels and involving hadrons with  non-zero spin.
   
 \item  It is important   to identify channels and energy regions that feature exotic hadrons and  can be reliably investigated with both lattice QCD and experiment. A detailed study of spectroscopic properties and structure on both sides could lead to conclusions that might  apply  more generally.  The lattice can straightforwardly and reliably study regions below  or  slightly above the lowest strong decay threshold for lattice QCD, but only few exotic hadrons  are  found  there. The  bound states $bb\bar u\bar d$ and $bb\bar u\bar s$ are firmly established on the lattice, but  it will be a formidable challenge to discover them in experiments. Lattice calculations also predict a supermultiplet of hybrid states, above strong decay thresholds, in both charmonium and bottomonium. The experimentally discovered states  $X(3872)$ and $cc\bar u\bar d$ lie near the thresholds and will render certain valuable conclusions from the lattice side, however their extreme closeness to the threshold makes  it hard to calculate their properties with a good precision. 

\item Turning to baryon systems, an ambitious but achievable goal for LQCD is to make reliable, robust predictions of
   the spectra and matrix elements of two-nucleon systems with full control of systematic uncertainties. These
   calculations and generalizations to three-body or larger systems will have profound impact on our knowledge of
   the structure of the lightest nuclei and provide inputs to nuclear effective theories with wide phenomenological
   application.

\item A longer-term challenge for the LQCD community is the determination of resonances decaying to three or more
  hadrons. This has seen significant investment of theoretical effort in recent years with promising early numerical
  results.   

 \item The determination of electroweak transitions of resonances is a relatively recent development but offers
   interesting new applications. Building on the rigorously extracted electromagnetic transition
   $\pi\gamma\rightarrow\rho\to\pi\pi$, a similar determination of transitions such as 
   $N\gamma\rightarrow\Delta\rightarrow N\pi$ and $K\gamma\rightarrow K^\ast\rightarrow K\pi$ can follow.
   A related, timely and relevant calculation that can be attempted in the medium term is a
   first look at transitions between stable hadrons and resonances via weak currents. These are significant in
   new physics searches, including for example
   the decay $B\rightarrow D^\ast l\bar{\nu}$ to $D^\ast\rightarrow D\pi$ which is relevant in
   violations of lepton flavor universality. Returning to spectroscopy, LQCD calculations of
   decay transitions can provide an alternative and complementary window on the nature and structure of resonances
   including exotics and further results can be expected within five years.
 \end{itemize}
 This white paper summarises progress and achievements to date in lattice hadronic physics and identifies relevant
 calculations that can be expected in the near future as well as suggesting a road map for
 further opportunities and describing the associated challenges for the field in the longer term. To realise the potential and impact of lattice hadronic physics requires continued access to computing resources and sustained investment in human capital.

 \section{Introduction}

	Quarks and gluons form a great variety of color-neutral hadrons~\cite{pdg2020}. With minimal valence quark content $\bar q_1q_2$ and $q_1q_2q_3$, so-called conventional hadrons (mesons and baryons, respectively) have long been studied. However, in the past two decades~\cite{pdg2020} candidates for non-conventional `exotic' hadrons with minimal valence quark content  $\bar q_1\bar q_2q_3q_4$ and $\bar q_1 q_2 q_3 q_4 q_5$ have been discovered in experiments. Other types of exotic hadron have also been postulated, such as hybrid hadrons which contain an excited gluonic field.
 
 The origin of these states lies within the strong interaction and at energy scales where non-perturbative effects dominate. Consequently, a perturbative expansion in the QCD gauge coupling is inapplicable. In the non-perturbative regime lattice QCD is widely used and possesses (in principle) systematically improvable systematic uncertainties. It is based directly on the QCD Lagrangian and spectroscopic  properties are determined from correlation functions. In turn these are calculated via 
the  numerical evaluation of the QCD path integral in a discretized and Euclidean space-time of finite extent. Practically all simulations employ the light dynamical  quarks, while most also employ dynamical strange quarks and some dynamical charm quarks. 
   
  In the following we discuss the status and prospects for using lattice QCD to study the spectroscopy of exotic and conventional hadrons.  The emphasis is on future prospects and this manuscript does not represent a full review of the existing results and the related literature. Instead it provides selected examples that illustrate the current status in each of the presented categories. The aim is to underline the challenges faced and to highlight possible directions for future studies using lattice and analytical methods. A related  manuscript with a similar aim is the USQCD white paper written in 2019~\cite{Detmold:2019ghl}. 
      
 Spectroscopic information for a hadron which is below, near, or above threshold  is commonly extracted from the discrete set of energies $E_n$ of QCD eigenstates $|n\rangle$ on a finite-volume lattice.
Two-point correlation functions are calculated on the lattice via numerical path integration, and the energies and overlaps $\langle O_j|n\rangle$ are then determined from the spectral decomposition
   \begin{equation}
\label{C}
C_{jk}(t)=\langle O_j (t) O_k^\dagger (0)\rangle= \sum_{n} \langle O_j|n\rangle  ~e^{-E_n t} \langle n|O^\dagger_k\rangle ~ . 
\end{equation} 
Here the interpolating operators $O_k^\dagger / O_j$ create/annihilate the hadron of interest, for example with a desired momentum $\vec{p}$ and combination of $J^P$ and flavor quantum numbers.\footnote{A discrete lattice with a finite extent has a reduced symmetry compared to an infinite-volume continuum. This means that instead of $J$ the relevant quantum number is an irreducible representation of the appropriate cubic point group -- this must be taken into account when computing correlation functions and in the analysis, but we will not discuss these details here.} 
When constructing the operators, care must be taken to ensure that they have appropriate structure(s) to effectively overlap with the eigenstates of interest. Furthermore, the study of a given high-lying hadron requires the extraction of all eigenstates with the same quantum numbers that lie within and below the energy region of interest.

We start with  the strongly stable hadrons and  then focus on hadrons that lie  above or slightly below the strong decay threshold. The electro-weak transitions of those are considered in Section 3, while methods to address their internal structure are discussed in Section 4.

\section{Hadron masses, strong decay widths and branching ratios } 
   
We will first discuss stable hadrons well below the relevant threshold for strong decay -- here there is a direct connection between energy eigenstates extracted from correlation functions and the hadron masses. For hadrons that can decay strongly or are near threshold, the connection is more involved and we will discuss these in later sections.

\subsection{Strongly stable hadrons}
\label{sec:stable_hadrons}

Among the vast number of hadrons, only a handful do not decay via the strong interaction~\cite{pdg2020}.  In the absence of electroweak interactions, the stable hadrons are the lowest-lying states of a given flavor content, $\pi$, $K$, $D$, $B^{(*)}$, $B_c^{(*)}$, $p$, $n$, $\Lambda$, $\Lambda_c$, $\Xi_{cc}$, etc.
The masses of these hadrons can be determined directly from the corresponding lattice ground state energy $m_0 = E_0(\vec{p}=\vec{0})$ from (\ref{C}) 
after it is extrapolated to zero lattice spacing ($a\!\to\! 0$), large volume ($L\!\to\! \infty$), and using or extrapolating to physical quark masses. Some more care is needed for hadrons that are close to threshold and we discuss these in Sec.~\ref{sec:bound-states}.  The masses of these hadrons have been accurately determined by a number  of lattice collaborations and are in close agreement with the experimentally observed values. Note that several lattice  studies of these hadrons include also the effects of isospin breaking and electromagnetism -- see Section~\ref{sec:isospinbreaking}.

Most calculations of charmonia ($\bar c c$) and bottomonia ($\bar b b$) use an approximation where the heavy quark and antiquark do not annihilate, and in this approximation there are a number of stable $\bar c c$ and $\bar b b$ hadrons.  Neglecting $\bar b b$ annihilation, the bottomonium system exhibits one of the richest spectra among strongly stable hadrons.   This spectrum is composed of several multiplets and has been extracted based on the energies (\ref{C})  within this approximation\footnote{  Arguments  concerning this approximation are presented in \cite{Lepage:1992,Bodwin:1995}. If in addition the  OZI-suppressed (disconnected) decays to $\bar b b$ and one or more light mesons are neglected, there are many states below the $B\bar B$ threshold which cannot decay strongly. Hybrids actually reside above $B\bar B$ threshold and their strong decays are omitted in \cite{Wurtz:2015mqa,Ryan:2020iog}.   } in \cite{Wurtz:2015mqa,Ryan:2020iog}. Many of these states, particularly the hybrids $\bar bGb$, are awaiting experimental discovery.   An alternative approach to study heavy quarkonia, which is not based on the direct computation of lattice energies, is to use static potentials and this is outlined in Sec. \ref{sec:potentials}.

\subsection{Generalities on hadrons that decay strongly or lie near thresholds}
\label{sec:scattering}

Most hadrons are in fact {\it hadronic resonances} that decay via the strong interaction. In particular, most, if not all, of the experimentally-discovered exotic hadrons are resonances or only slightly below the threshold for strong decay. Such hadron resonances are not asymptotic states of QCD and this represents the main challenge for a rigorous theoretical study of their properties.

Resonances correspond to pole singularities of the corresponding scattering matrices, $T(E_{cm})$, in the complex $E_{cm}$ plane, where $E_{cm}$ is the center-of-momentum energy. The mass $m_R$ and width $\Gamma$ of the resonance are related to the position of the pole, $E^p_{cm}=m_R \pm i\frac{1}{2}\Gamma$, and its couplings to the various channels are related to the residues at the pole. In general, $T$ is a matrix in the space of all the relevant coupled channels (coupled partial waves and/or hadron-hadron channels). For elastic scattering it simplifies to a $1 \times 1$ matrix. For a narrow, isolated resonance in elastic scattering, $m_R$ and $\Gamma$ correspond to the mass and width in a Breit-Wigner parametrization of $T(E_{cm})$ for real energies.

 A {\it strongly stable state}, i.e.\ a {\it bound state}, corresponds to a pole in $T$ at a real $E_{cm}$ below threshold with $p=i|p|$, where $p$ denotes the momentum of each hadron in the center-of-momentum frame. The mass of the hadron corresponds to the pole position, $E_{cm}^p=m$. A near-threshold bound state can be affected by the threshold, and this can be taken into account by considering low-energy scattering and extracting $T(E_{cm})$. If instead the pole occurs for $p=-i|p|$, there is a {\it virtual bound state} rather than a bound state. It should be noted though, that the former is a new, {\it i.e.} asymptotic state while the latter is not, {\it i.e.} it is an element of the two-particle continuum.

Below we discuss the status, challenges and prospects for determining  scattering matrices $T(E_{cm})$ in various quantum number channels using lattice QCD.  The information gained gives insight into $T$ for real energies above and somewhat below the threshold.

\subsection{Extracting scattering matrices from lattice QCD}

The most widely applied and rigorous method to extract the scattering matrix from ab-initio lattice QCD simulations is the so-called {\it L\"uscher formalism} \cite{Luscher:1986pf,Luscher:1990ux} together with  its generalizations (e.g.\ \cite{Rummukainen:1995vs,Kim:2005gf,Lage:2009zv,Hansen:2012bj,Gockeler:2012yj,Briceno:2014oea,Briceno:2015csa,Briceno:2017tce}), which are based on finite volume quantization conditions.
The method rigorously relates the infinite volume scattering matrix $T(E_{cm})$ and the discrete eigenenergies $E_{cm}$ of the system in a finite volume. The latter can be obtained from lattice simulations via the previously mentioned correlators (\ref{C}). The method is applicable to hadron-hadron scattering with an arbitrary set of coupled hadron-hadron channels and partial waves, involving hadrons with zero or non-zero spin, below the production threshold for three or more hadrons.
In the following we sketch the essential ideas of the formalism that enable an understanding of its applications and challenges. An illuminating quantum field theory based derivation is presented in \cite{Kim:2005gf} and numerous applications are reviewed in, for example, \cite{Briceno:2017max,Brambilla:2019esw,Padmanath:2019wid}.

The L\"uscher quantisation condition can be written as $\mathrm{det}[F^{-1}(E_{cm})-8\pi i T(E_{cm})]=0$, where $F^{-1}(E_{cm})$ is a matrix of known functions, and both $F^{-1}$ and $T$ are matrices in the space of coupled hadron-hadron channels and partial waves.\footnote{The reduced symmetry of the finite-volume lattice introduces some additional mixing of partial waves, but we will not discuss this here -- see e.g.~\cite{Briceno:2017max}.}
The solutions of this determinant equation   are the finite-volume energies $\{E_n\}$ in the relevant quantum number channel, and hence the $\{E_n\}$ can be determined if the scattering matrix $T(E_{cm})$ is known. However, to study resonances using lattice QCD, the inverse problem must be solved.
For elastic scattering, $T(E_{cm})$ is a $1\! \times \! 1$ matrix, and there is a one-to-one relation between $E_n$ and $T(E_{n})$.\footnote{For a system with total momentum $\vec P$, the  $E_n$ renders $T(E_{cm})$ at $E_{cm}=(E_n^2-\vec P^2)^{1/2}$.} If many finite-volume energies can be extracted over the energy region of interest, the energy dependence of $T$ can be mapped out. Analytically continuing a parameterisation of $T(E_{cm})$ to the complex $E_{cm}$ plane, the presence of any resonances and bound states can be determined.
On the other hand, for inelastic scattering $E_n$ provides a constraint on $T(E_n)$, but this is an undetermined problem. Unless some trick can be used to achieve many constraints at a given $E$, the energy dependence of $T$ must be parameterised in an appropriate way and the parameters varied so as to best describe the finite-volume energies. Any poles in the complex energy plane can then be found and the presence of resonances  determined.

Below we discuss various types of scattering systems, beginning with the simplest case of elastic scattering of spin-less hadrons and then moving to more complicated problems involving hadrons with non-zero spin and/or coupled channels. Scattering involving three (or more) hadrons is more challenging -- significant progress has been made both in the extension of the formalism and in first lattice QCD calculations.

Currently, most lattice QCD studies of scattering do not quantify all the various systematic uncertainties, unlike the calculations of stable hadrons described above. The majority are performed at a single lattice spacing.\footnote{Studies with a number of lattice spacings are briefly discussed in Sec.~\ref{sec:improvements}.} Furthermore, they omit $\bar cc$ or $\bar bb$ annihilation, as well as isospin breaking effects and electromagnetic interactions. With the notable exception of $\pi\pi$ and $K\pi$ scattering, many employ one or only a few values of the light quark masses $m_{u/d}$, which typically are heavier than the physical values. This often significantly simplifies a study because the number of open channels is reduced and certain channels with three or more hadrons are closed. In addition, lattice energies have smaller statistical errors at larger-than-physical light quark masses due to the signal-to-noise behaviour of lattice correlators. The partial width $\Gamma$ of a resonance to a decay channel depends on the available phase space and is therefore sensitive to the masses of the hadrons and hence the quarks. When comparing calculations with different quark masses and experimental results, it is therefore convenient to compare a coupling $g$ where the phase space has been divided out, $\Gamma= g^2 ~p^{2l+1}/E_{cm}^2$, rather than the width itself.
It is important to note that performing calculations with various unphysical light-quark masses can be a useful tool for discerning the structure of a hadron, as we discuss below.

\subsection{Resonances decaying only to one pair of spin-less hadrons}\label{sec:one-channel}
 
 We start by focusing on resonances that solely or predominantly decay to only one pair of spin-less hadrons. There are only a few  of this sort in nature, however, these resonances are the easiest to rigorously study. 
 Of these the reasonably-narrow vector resonances  $\rho \to \pi\pi$ and $K^*\to K\pi$, sometimes referred to as ``vanilla'' resonances, are the only ones that have been considered by a number of lattice simulations.  The resulting   masses and widths/couplings  are in good agreement with experiment.  These   therefore represent a (mostly) solved problem.
 
Scalar resonances $\sigma \to \pi\pi$ \cite{Briceno:2016mjc,Guo:2018zss} and  $\kappa\to K\pi$  \cite{Brett:2018jqw,Rendon:2020rtw, Wilson:2019wfr} are so broad that they appear very different from ``vanilla'' resonances and the scattering phase shift does not show an obvious rise through $90^\circ$.  
So far only a few values of the   light quark mass   have been studied. Nevertheless, the poles seem to be roughly approaching those extracted in experiment \cite{pdg2020}  as   the light quark mass is lowered. Future simulations closer to or nearly at the physical quark mass values would be of a great interest here.   It will be valuable to robustly determine the pole position based on the Roy equation formalism \cite{Caprini:2005zr,Descotes-Genon:2006sdr},  which has been used for this purpose on the experimental data.  
 
The scalar resonance $D_0^*$ has represented a puzzle since its mass  $m\simeq 2.34~$GeV in the PDG \cite{pdg2020} is reported to be almost degenerate to the  scalar $D_{s0}^*$. Interestingly, lattice simulations of $D_0^*\to D\pi$ \cite{Gayer:2021xzv,Moir:2016srx,Mohler:2012na}%
\footnote{The BW fit at real energies rendered $m_{D_0^*}\simeq 2.32(2)~$GeV \cite{Mohler:2012na}, while the corresponding pole in the complex plane based on the same lattice data is at $m_{D_0^*}\simeq 2.12(3)~$GeV (the pole was not  quoted in \cite{Mohler:2012na}).}  find a pole at a lower mass $m\simeq 2.1-2.2~$GeV, consistent with suggestions~\cite{vanBeveren:2003kd,vanBeveren:2006st,Du:2020pui} and reanalysis of the experimental data \cite{Du:2020pui}, and lower in mass than the $D_{s0}^*$. This makes the $D_0^*$ a more natural partner of  $D_{s0}^*$. These two states complete the SU(3) flavor-triplet  according to the HQET+ChPT approach  \cite{Kolomeitsev:2003ac,Du:2017zvv,Albaladejo:2016lbb,Guo:2018tjx}. The same approach   suggests also the existence of a flavor sextet of scalar 4-quark states some of  which should feature in the coupled channel scattering  discussed in Section \ref{sec:coupled}, consistent with lattice studies of isospin-0 $D\bar{K}$ scattering~\cite{Cheung:2020mql}. Simulations of scalar resonances $D_0^* $, $B_0^*$ at multiple light quark masses close to the physical point, along with calculations with heavier quark masses and SU(3) flavor symmetry, would be valuable to reinforce this emerging picture and determine if it persists with physical-mass light quarks. In addition, accurate values of the scattering lengths that enter as an input to the low energy constants of the HQET+ChPT  approach \cite{Du:2017zvv,Albaladejo:2016lbb} would be valuable as a function the quark masses.  An update would be very desirable as currently the input comes from an impressive but rather old simulation \cite{Liu:2012zya}. 
   
  The conventional charmonium resonances $\psi(3770)$ with $J^{PC}=1^{--}$ and $X(3842)$ with $3^{--}$  were  extracted from $D\bar D$ scattering \cite{Piemonte:2019cbi} and the results  compare  reasonably to experiment   \cite{LHCb:2019lnr}.

  \subsection{Near-threshold bound states and doubly heavy tetraquarks }\label{sec:bound-states}
  
A stable hadron well below threshold can be directly related to a computed lattice energy, as discussed in Sec.~\ref{sec:stable_hadrons}. However, when a hadron is close to threshold, finite-volume effects can be significant and so some care is needed to determine whether there is a bound state, a resonance or just an attractive interaction. In this case, the scattering matrix needs to be extracted, the pole singularities determined and hence whether a bound state (or resonance) is present can be deduced, see Sec.~\ref{sec:scattering}. A strongly stable state with a mass $m$    corresponds to a pole in the  scattering matrix $T(E_{cm})$ at $E_{cm}=m$.
An interesting question is the nature of a bound state just below $H_1H_2$ threshold: whether it can be considered (predominately) a $H_1H_2$ molecule, a compact tetraquark, or a $\bar qq$ bound state whose formation in QCD is only weakly affected by the $H_1H_2$ interactions. In other words, what are the dominant Fock components.
There exist certain criteria, based on measured observables, which allow one to distinguish these alternatives from each other.
As before, the simulation of scattering is more amenable if both hadrons are spin-less and more difficult if one or both carry spin.

The bound states $X(3872)$ in $D\bar D^*$ scattering \cite{Prelovsek:2013cra,Padmanath:2015era} and $D_{s0}^*$ in $DK$ scattering  \cite{Mohler:2013rwa,Bali:2017pdv,Cheung:2020mql}  were the first hadronic bound states identified in this way in lattice QCD calculations. It was found that threshold effects are important and that the masses of both states are pushed down compared to expectations of $\bar q q$ states in quark models, thereby putting them in closer agreement with experiment. It is believed that these two states   have significant molecular/tetraquark Fock components. Hence an  open question  is whether there exist additional nearby states  with the same quantum numbers that would be dominated by conventional $\bar qq$ Fock components.  To date, neither experiments nor lattice studies have found such states.   The existing lattice studies of $X(3872)$ \cite{Prelovsek:2013cra,Padmanath:2015era}  call for several challenging improvements: accounting for the  isospin breaking effects, the presence of multi-hadron decay channels $J/\psi \pi\pi$ and  $J/\psi \pi\pi\pi$, for the dynamical mixing of partial waves $l=0,2$ in the $D\bar D^*$ scattering with $J^P=1^+$ and simulations on (much) larger volumes. On the other hand, the simulations of the ``vanilla'' bound state $D_{s0}^*$ in the mesonic sector have matured and don't rely on any major simplifications. The bottom partner  $B_{s0}^*$ was found as a bound state  in $BK$ scattering \cite{Lang:2015hza}. This state still awaits experimental confirmation. 
 
Turning towards open bottom tetraquarks with flavors $bb\bar u \bar d$,   $bb\bar u \bar s$   and $J^P=1^+$, we note these are   among the most convincing exotic hadrons that have been reliably established on the lattice \cite{Bicudo:2015vta,Francis:2016hui,Hudspith:2020tdf,Leskovec:2019ioa,Junnarkar:2018twb} as well as with the phenomenological methods, e.g. \cite{Karliner:2017qjm, Eichten:2017ffp}.  Their masses were found to lie significantly below the strong decay thresholds. As a result the determination of their masses  is relatively straightforward.
Their binding increases with increasing $m_b$ and decreasing $m_{u/d}$ \cite{Junnarkar:2018twb,Francis:2018jyb}. This is expected for states dominated by diquark-antidiquark Fock component. The only study aimed to extract the scattering amplitude for doubly-bottom tetraquarks near threshold  considered    $BB^*$ scattering \cite{Leskovec:2019ioa}.  The analogous study would be valuable for coupled-channels   $BB_s^*-B_sB^*$. The prospect for their discovery in existing experiments via exclusive decays is complicated, while there are certain prospects for inclusive detection where they would appear together with doubly bottom baryons \cite{Gershon:2018gda}. 
Studies of these channels with static (infinitely heavy) $b$ quarks are discussed in Sec.~\ref{sec:potentials}.

The doubly-charmed tetraquark $T_{cc}=cc\bar d\bar u$  with a likely assignment  $J^P=1^+$ and $I=0$ has been discovered $0.4~$MeV  below $DD^*$ threshold in an impressive study by LHCb in 2021 \cite{LHCb:2021vvq,LHCb:2021auc}. On the lattice the ground-state finite-volume energy for $cc\bar u\bar d$  with $I(J^P)\!=\!0(1^+)$ has been found slightly below threshold ($\simeq 20~$MeV) \cite{Junnarkar:2018twb}, while it was found close to or slightly above the threshold in \cite{Cheung:2017tnt}. This information alone does not resolve the question of whether these channels feature a bound state, a virtual bound state, a resonance or just a weakly interacting pair of hadrons that is not accompanied by any tetraquark state.  The only way to establish the existence of a state near threshold is to determine the scattering amplitude. The   $DD^*$ scattering amplitude was recently extracted in \cite{Padmanath:2022cvl}, where   $T_{cc}$ is found as a virtual bound state $\simeq\! 10~$MeV below threshold   at $m_\pi \simeq 280~$MeV.  As  $m_\pi$ is decreased towards the physical, it is expected that the virtual bound state would approach the threshold  and eventually turn to the bound state  \cite{Padmanath:2022cvl}. This has to be verified using the actual lattice studies in the near future.

Other candidates  for doubly-heavy mesons $Q_1Q_2\bar q_3\bar q_4$  ($Q=c,b,~q=u,d,s$) are expected to lie near or above strong decay thresholds \cite{Hudspith:2020tdf,Junnarkar:2018twb,Cheung:2017tnt}.  The ground-state finite-volume energy  for $bc\bar u\bar d$  was found slightly above threshold at 
$m_\pi\simeq 200~$MeV   \cite{Hudspith:2020tdf}, while it was found   slightly below it 
for range of $m_\pi\geq 480~$MeV \cite{Padmanath:2021qje}. This does not yet resolve the question of whether these channels feature an exotic hadron.    The way to establish the existence of a state near or above threshold is to determine the scattering amplitude, which has not been done yet (with exception of $BB^*$ \cite{Leskovec:2019ioa} and $DD^*$ \cite{Padmanath:2022cvl} studies mentioned above).    Further studies of scattering matrices    for the one-channel or two-channel scattering   $(Q_1\bar q)(Q_2\bar q)$ near threshold are feasible in the forthcoming years, since the tetraquarks $Q_1Q_2\bar q_3\bar q_4$ are the lowest lying states  with the given quantum numbers.
 Note that $\bar Q_1Q_2\bar q_3q_4$ are more challenging since they can decay to a heavy quarkonium and a light meson.  
 
 It would be valuable to determine the quark mass dependence of the pole positions  and how these depend on the positions of the thresholds. This information could give hints towards the nature of the states.  It seems conceivable to determine their masses   with  an accuracy of $\mathcal{O}(10~\rm{MeV})$ at the currently simulated quark masses. However an  accuracy of $\mathcal{O}(1~\rm{MeV})$ on the mass of  $T_{cc}=cc\bar u \bar d$ at the physical quark masses is not feasible at this time as further effects, e.g. isospin breaking have to be included.
 
It will be important   to explore  why $X(3872)\simeq \bar cq\bar qc,\bar cc$ and $T_{cc}\simeq cc\bar u \bar d$ with $J^P=1^+$ are both situated within $1~$MeV of the $D\bar D^*/ DD^*$ thresholds. Certain mechanisms that may be responsible for their existence are fundamentally different.   The $X(3872)$ is likely affected by the conventional $\bar cc$ Fock component, while $T_{cc}$ is likely affected by the Fock component $[cc]_{\bar 3_c}[ud]_{3_c}$ that binds the so-far unobserved $T_{bb}$.  

   It will be valuable to  establish whether any $\bar cc\bar cc$ bound states feature in the scattering of two charmonia since  the lower-lying partners of experimentally discovered $X(6900)\simeq \bar cc\bar cc$   are expected in certain models \cite{Giron:2020wpx,Bedolla:2019zwg,Dong:2020nwy}. Possible bound states in  $\bar bb\bar bb$ have already been searched for in $\eta_b\eta_b$, $\Upsilon\Upsilon$, $\eta_b\Upsilon$ scattering by HPQCD and no bound state was found \cite{Hughes:2017xie}. 
 
Lattice studies of di-baryon bound states and resonances are discussed in Section \ref{sec:dibaryons}.

  \subsection{Resonances  decaying to several two-hadron  states or to particles with spin }\label{sec:coupled}
  
  Most conventional and exotic resonances can decay strongly to several different final states. The rigorous lattice study of these resonances is considerably more difficult. Nevertheless, some progress has been made in recent years and the Hadron Spectrum Collaboration has managed to determine the masses, widths and branching ratios of resonances in a variety of quantum channels, for example~\cite{Dudek:2014qha,Wilson:2014cna,Wilson:2015dqa,Dudek:2016cru,Moir:2016srx,Briceno:2017qmb,Woss:2018irj,Woss:2019hse,Woss:2020ayi}. However, a number of interesting  resonances have not been rigorously studied yet --  this applies to most of the experimentally discovered exotic resonances, most of the resonances with heavy quarks and those that lie high above threshold. This is a topic where major effort should be made in the near-future studies.   \\ 
     
{\noindent \bf Scattering of particles with spin in a single hadron-hadron channel}

 \vspace{0.1cm}
  
When one or both hadrons have non-zero spin, several combinations of total spin, $\vec s=\vec s_1+\vec s_2$, and orbital angular momentum, $\vec l$, can dynamically mix to give the same $J^P$. This leads to a non-diagonal $T_{ll^\prime}$.  Such a scattering matrix has been successfully extracted for $\rho\pi$ scattering in the repulsive channel with $J^P=1^+$ and $I=2$, where partial waves $l=0$ and $l=2$ dynamically mix \cite{Woss:2018irj}. This was done for heavy $u/d$ quarks such that the $\rho$ is stable.  $N\pi$ scattering in $l=1$ and $I=3/2$ leads to the $\Delta$ resonance \cite{Andersen:2017una,Silvi:2021uya}, while incorporation of $N\pi\pi$ decay mode is awaiting the future. The analogous channel with $I=1/2$ did not render a low-lying Roper resonance when treating $N\pi$ and $N\pi\pi$ channels as decoupled \cite{Lang:2016hnn,Wu:2017qve}. \\

{\noindent \bf Coupled-channel scattering of spin-less hadrons} 

 \vspace{0.1cm}
    
    Coupled-channel scattering of spin-less hadrons has been explored mostly by the Hadron Spectrum Collaboration. Complete multiplets of resonances with $J=0,1,2$ composed of $u,d$ and $s$ quarks have been obtained. For example, $J=0$ strange resonances were extracted from coupled  $K\pi-K\eta$ scattering in a pioneering study \cite{Dudek:2014qha,Wilson:2014cna}, while $J=0$ isoscalar resonances were extracted from coupled  $\pi\pi-K\bar K-\eta\eta$ scattering \cite{Briceno:2017qmb} where $f_0(980)$ manifests as a dip in   $\pi\pi$ rate near $K\bar K$ threshold. 
    
    Coupled-channel scattering is much less explored in the heavy-quark sector and this is where  significant effort has to be made in the future. Coupled $D\pi-D\eta-D_s \bar K$ scattering with $m_\pi\simeq 390~$MeV \cite{Moir:2016srx} led to a scalar $D_0^*$ meson lighter than the $D_{s0}^*$, rendering them more natural members of the same flavor triplet (see related   discussions on the $D_0^*$ and $D_{s0}^*$ in Sections \ref{sec:one-channel} and \ref{sec:bound-states}). The reanalysis \cite{Albaladejo:2016lbb} of the same lattice energy levels leads to the suggestion of an extra pole which which would be a member of a heavier $SU(3)$ flavor sextet~\cite{Kolomeitsev:2003ac,Du:2017zvv,Albaladejo:2016lbb}.  The eigen-energies from a lattice simulation at SU(3) symmetric point    suggest a bound state or a virtual bound state in flavor sextet  and repulsion in flavor 15-plet \cite{Gregory:2021rgy}.  
    Future lattice simulations should explore  
the scalar charmed mesons at lighter $m_{u/d}$ and the corresponding beauty mesons. 
  
  Coupled $D\bar D-D_s\bar D_s$ scattering with $I=0$  rendered the expected charmonium resonances 
  $\chi_{c0}(2P)$ and  $\chi_{c2}(2P)$, as well as  two  unconventional scalar states just below $D\bar D$ and $D_s\bar D_s$ thresholds \cite{Prelovsek:2020eiw}. The $D \bar D$ bound state was not claimed by experiment, but predicted in \cite{Gamermann:2006nm} and found in the analysis of experimental data for example in \cite{Deineka:2021aeu}. The state slightly below $D_s\bar D_s$ shares similar features as experimental $X(3915)$ as it lies just below $D_s \bar D_s$ and has a very small/negligible coupling to $D\bar D$.  
  Future lattice simulations should aim to go beyond several simplifications of this study listed in Section 5 of \cite{Prelovsek:2020eiw}.  \\

{\noindent \bf Coupled-channel scattering of hadrons with non-zero spin} 

 \vspace{0.1cm}
     
Calculations of coupled-channel scattering involving hadrons with non-zero spin are needed to address most of the exotic   resonances observed in the experiments. However, due to severe challenges involved in doing this, such a scattering matrix has been determined using the L\"uscher method only for a few channels that involve just light and strange quarks.  The $b_1$ resonance with a mass $1.38~$GeV  has been extracted from coupled $\pi\omega-\pi\phi$ scattering at $m_\pi\simeq 390~$MeV~\cite{Woss:2019hse}. 
The same collaboration found a candidate for a hybrid meson $\pi_1$ resonance with the exotic quantum numbers $J^{PC}=1^{-+}$ by studying eight coupled channels in the $SU(3)$ flavor symmetry limit at $m_\pi\simeq 700~$MeV~\cite{Woss:2020ayi} -- using a crude extrapolation to physical light-quarks, the resonance is found to couple dominantly to $b_1\pi$ and is likely related to $\pi_1(1564)$ \cite{JPAC:2018zyd}. For the same $m_\pi\simeq 700~$MeV, light-meson resonances with $J^{PC}=1^{--},2^{--},3^{--}$ were found in pseudoscalar vector scattering~\cite{Johnson:2020ilc}. Using unphysically-heavy light quarks in these calculations meant that the resonances do not have any three-hadron decay modes and the scattering vectors ($\rho,b_1,..$) are strongly stable.  Analogous studies at the physical quark masses will be much  more difficult due to decay modes involving three or more hadrons and also due to the statistical uncertainties on energies.
   
   In systems with heavy quarks, the coupled-channel scattering matrices involving hadrons with non-zero spin have not been extracted yet with the L\"uscher's method. Several studies computed the eigen-energies for certain quantum numbers of interest, but the   number of energy levels  was not large enough to constrain the energy dependence of scattering matrices.  
 The determination of these scattering matrices should be one of the important aims of the near-future lattice studies, in particular for the exotic resonances that have been discovered in experiments.

    The exotic resonance $Z_{c}(3900)\simeq \bar cc\bar du$   appears in coupled $J/\psi \pi-D\bar D^*-\eta_c\rho$ scattering and certain eigen-energies have been determined in \cite{Prelovsek:2014swa,Cheung:2017tnt,Chen:2014afa}. The coupled channel scattering matrix, however, has been determined only using  the HALQCD approach \cite{HALQCD:2016ofq},  which suggests that the large coupling between $J/\psi \pi-D\bar D^*$ channels    is responsible for the existence of $Z_c(3900)$.  The near-future lattice simulations should extract this scattering matrix up to energies of $4~$GeV, which seems a conceivable goal if $\rho$ is treated as stable.  

  A number of pentaquarks $P_c\simeq \bar cc uud$ have been discovered in the $J/\psi p$ channel by LHCb \cite{LHCb:2015yax,LHCb:2019kea}. Only one simulation   explored the  energies   where $P_c$ reside -- it considered  $J/\psi p$ scattering in a one-channel approximation \cite{Skerbis:2018lew}. It identified a large number of   near-degenerate $J/\psi p$ eigenstates, which arise since both the nucleon and $J/\psi$ carry spin. The results show that  $J/\psi p$ scattering alone does not render  $P_c$ resonances, which indicates that the coupling to other channels (most likely $cqq+\bar cq$) must be responsible for their existence in experiment \cite{Skerbis:2018lew}. Future lattice simulations should aim to incorporate also at least the $\Sigma_c^+\bar D^0$ and $\Sigma_c^+\bar D^{0*}$ channels, since three  observed pentaquarks lie 
  near these thresholds and  may be likely dominated by the molecular Fock components. Rigorous extraction of the scattering matrices for all open channels    $cqq+\bar cq$ and $\bar cc +uud$ does not seem feasible in the near future, therefore it will be valuable to 
  undertake studies to determine if certain channels could be neglected and how to address this problem in practice. 
  
        A very interesting resonance  $X(6900)\simeq \bar cc\bar cc$ was discovered in $J/\psi J/\psi$ decay by LHCb. It lies significantly above the lowest strong decay threshold    $\eta_c\eta_c$ and can decay to a number of final states, therefore a rigorous study  seems difficult. However, a simulation of coupled $\eta_c\eta_c -J/\psi J/\psi$ channel scattering 
        at lower energies closer to the threshold seems more feasible and could find lower-lying  $\bar cc\bar cc$ tetraquarks which are predicted by certain models \cite{Giron:2020wpx,Bedolla:2019zwg,Dong:2020nwy}. Simulation of $\eta_b\eta_b -\Upsilon \Upsilon$ scattering would also be of interests, while $\bar bb\bar bb$ tetraquark candidates  are still awaiting experimental discovery. 
        
        In general, significant progress can be expected for  resonances that decay only to a few two-hadron final states and do not         
        lie very high above the lowest threshold.           On the other hand, resonances that do not meet these criteria,  will likely remain an unsolved problem within lattice simulations for some time, and some of them are mentioned in Section \ref{sec:unsolved}.

\subsection{Isospin breaking and electromagnetic effects}
\label{sec:isospinbreaking}

Taking into account isospin breaking and QED effects introduces an extra level of complication for lattice QCD calculations. For this reason they are typically omitted, in particular when their role is expected to be subdominant. As a result only very few lattice studies of scattering where bound states or resonances appear have included them.
An example where they have been included is in the recent study of multi-channel systems   where bound states or resonances do not appear \cite{NPLQCD:2020ozd}. 
Nevertheless, there are further cases where it would be valuable to take it them into account. In particular $X(3872)\simeq \bar cc,~\bar cq\bar qc$ is such a case as it has comparable branching ratios for the decays to $J/\psi \rho$ and $J/\psi \omega$ that carry different isospin. Such a study would be more challenging than \cite{Prelovsek:2013cra,Padmanath:2015era}, since the dominant channel $D\bar D^*$  will be replaced by two  coupled channels $D^0\bar D^{*0}$ and $D^-\bar D^{*+}$ in the simulation. The observed $X(3872)$ is located within $1~$MeV of   $D^0\bar D^{*0}$  threshold and $8~$MeV below $D^-\bar D^{*+}$ threshold.  For the newly discovered $T_{cc}$ with a binding energy of $0.4~$MeV, isospin breaking and QED effects could also play an important role.

\subsection{Static potentials for systems with two heavy quarks }\label{sec:potentials}

Hadronic systems that contain two heavy quarks, i.e. $QQ$ or $\bar QQ$ with $Q=b$, and additional light degrees of freedom  (gluons $G$ and/or light quarks $q=u,d,s$), can be addressed via the Born-Oppenheimer approximation. This is because there is an effective separation of scales between the velocity of the heavy quarks compared to those of the other particles. For such systems lattice studies often consider simulations with infinitely heavy, static quarks that are separated by  a spatial distance $r$. The eigenenergies of these systems as a function of   $r$  can then be connected to static potentials $V(r)$. Once the potential is extracted via fit ansatz to the discrete lattice data the motion of the heavy degrees of freedom is studied in this fitted potential. For example, one can determine the masses of the bound states or resonances via the Schr\"odinger equation. 
 
In this non-relativistic picture, the static $\bar QQ$ potential exhibits Coulomb attraction at short distances and a linear rise related to confinement at large distances. Large energies contained in the gluon string at large separations allow for the creation of $\bar qq$ pairs. This results in the breaking of the gluon string to $B\bar B$  \cite{Bali:2005fu} and   $B_s \bar B_s$ \cite{Bulava:2019iut}. Future simulations should explore this also at lighter $m_{u/d}$, smaller $a$ and determine also the potential at small separations $r$.
Such studies could give information on the bound states and resonances with isospin zero near the $B\bar B$  threshold \cite{Bicudo:2019ymo}, which is currently based on the potentials of \cite{Bali:2005fu}.

Excited state potentials which are related to the hybrids $\bar QGQ$  have been mostly studied in quenched QCD \cite{Juge:1999ie,Schlosser:2021wnr} as this avoids their strong decay.  Determining these potentials at smaller $a$ and $r$ was  partly accomplished  in \cite{Schlosser:2021wnr}.  The potentials permit the estimation of the masses for hybrid states $\bar b Gb$ \cite{Guo:2008yz,Brambilla:2018pyn,Brambilla:2019jfi}, these in turn can be compared to the masses obtained from lattice calculations that use relativistic $b$ quarks in a dynamical setup. An example of such a lattice simulation is \cite{Ryan:2020iog} where the strong decays are omitted.  The spectrum  of heavy hybrids in  dynamical QCD that  incorporates their strong decays,  is a challenge left for future analytical studies and simulations. An encouraging pioneering study of $\pi_1$ in the light sector has been reported in \cite{Woss:2020ayi}. All $\bar cGc$ and $\bar bGb$ with exotic $J^{PC}$ are still awaiting experimental discovery. 
 
 First lattice evidence for the existence of a  strongly stable $b b \bar u \bar d$ tetraquark with $J^P = 1^+$ and a $b b \bar u \bar d$ tetraquark resonance with $J^P = 1^-$ was found based on the attractive static potentials   \cite{Bicudo:2012qt,Bicudo:2015vta,Bicudo:2015kna,Bicudo:2016ooe,Bicudo:2017szl}. Simulations with non-static bottom quarks (see Sec. \ref{sec:bound-states}) confirmed the existence of the stable state with $J^P = 1^+$, while the resonance with $J^P = 1^-$ has not yet been studied in such a framework. 
   Static potentials for other doubly-heavy channels did not give an indication for bound states or near-threshold resonances, e.g. \cite{Bicudo:2015vta}.   It is not surprising that such an approach  does not permit predictions for the observed $T_{cc}$ \cite{LHCb:2021vvq} as the Born-Oppenheimer approximation is not reliable away from the very heavy, deeply non-relativistic regime.  
    
 Considering exotic $Z_b\simeq \bar bb\bar du$ resonances  discovered by Belle \cite{Belle:2011aa}, the main challenge for their  lattice study is that they strongly decay to $\bar bu+\bar db$ as well as to numerous final states $\bar b b+\bar d u$. This was taken into account  in the simulations \cite{Peters:2016wjm,Prelovsek:2019ywc,Sadl:2021bme}, which suggest that there is significant attraction between $B$ and $\bar B^*$ mesons which is in turn responsible for the existence of $Z_b$.  Analogous simulations of the $\bar bb\bar su$ and $
  \bar bbq_1q_2q_3$ channels will be valuable for the study of the  beauty partners $Z_{bs},~ P_b,~P_{bs}$, i.e. states  that have been experimentally discovered in the charmonium-like sector. Some general analytic considerations are provided in \cite{Braaten:2014qka,Soto:2020xpm,Brambilla:2021mpo}. A number of future improvements are necessary to study such closed-bottom states beyond the simplifications used in  \cite{Peters:2016wjm,Prelovsek:2019ywc,Sadl:2021bme}.   The potential at very small $r$  is needed from the analytical side, the mixing between the static quantum channels within   a given $J^P$   needs to be taken into account and the decays of light resonances within $\bar b b+\bar q_1 q_2$ need to be incorporated. 

The ground state static potential of $\bar QQ$ is found to be only very mildly affected by the presence of various light hadrons at rest  \cite{Alberti:2016dru}. Energy shifts of at most a few MeV were found in this lattice investigation  of the hadroquarkonium picture.

 \subsection{Di-baryon and multi-nucleon systems}\label{sec:dibaryons}

The two-baryon sector exhibits rich dynamics in nature including a two-nucleon bound state, the deuteron. It has long been conjectured to include also a possible $SU(3)$  flavor-singlet  bound state  $udsuds$, i.e. the $H$ dibaryon below $\Lambda\Lambda$ threshold.   Pioneering LQCD studies by several groups have used L{\"u}scher's method to obtain nucleon-nucleon, as well as nucleon-hyperon and hyperon-hyperon, scattering amplitudes using larger-than-physical quark masses and have found evidence for the existence of the deuteron  and other two-baryon bound states, see Refs.~\cite{Beane:2010em,Drischler:2019xuo,Davoudi:2020ngi} for reviews.
These results can be matched to nuclear EFT and model results in order to fix parameters describing baryon-baryon interactions in the (hyper-)nuclear Hamiltonian.
LQCD calculations of two-nucleon systems with realistically achievable precision will provide validation of LQCD methods and experimentally inaccessible but physically interesting information about the quark-mass dependence of nuclear observables~\cite{Epelbaum:2013wla,Beane:2013kca}. 
Predictions of hyperon-nucleon and hyperon-hyperon interactions will provide important inputs to determinations of the equation of state of dense matter and neutron star structure~\cite{Beane:2012ey}.
 
LQCD calculations of multi-nucleon systems have been  performed using larger-than-physical quark masses in order to lessen the severity of the signal-to-noise problem affecting multi-baryon correlations functions. Results have been used to develop methods for matching between LQCD and effective theories as well as to understand the quark-mass dependence of multi-baryon observables.
In particular, pionless EFT has been matched to LQCD results~\cite{Beane:2012vq} for the binding energies of $B=2-4$ nuclei with $m_\pi = 806$ MeV and found to provide accurate descriptions of few-nucleon systems by validating EFT predictions with LQCD results~\cite{Barnea:2013uqa,Eliyahu:2019nkz,Detmold:2021oro}. The EFT was subsequently used to make predictions for the binding energies of nuclei as large as ${}^{16}$O and ${}^{40}$Ca with the same quark mass values~\cite{Contessi:2017rww,Bansal:2017pwn}, demonstrating that EFT can be used to extend LQCD results to larger nuclei. Robust LQCD predictions of the energy spectra and matrix elements of two-nucleon systems using physical quark masses and with controlled systematic uncertainties will provide  insight both on the emergence of nuclear structure from the underlying dynamics of QCD, as well as precise constraints on effective theories with applications ranging from nuclear astrophysics to new physics searches using nuclei as targets.

The extracted eigen-energies and scattering amplitudes  of two-nucleon  and $H-$dibaryon  systems    
still depend on the choice of the   interpolators $O_i$, on whether all elements of the correlation matrix $C_{ij}$ (\ref{C}) are  evaluated and on the lattice spacing \cite{Detmold:2012eu,Doi:2012xd,Francis:2018qch,Horz:2020zvv, Amarasinghe:2021lqa,Green:2021qol,  Beane:2012vq,Beane:2013br,Orginos:2015aya,Wagman:2017tmp,Illa:2020nsi,Yamazaki:2012hi,Yamazaki:2015asa,Berkowitz:2015eaa}.  
Future simulations will benefit by  calculating all elements of the correlation matrix and extracting the eigen-energies via the variational analysis. This approach has  been widely and   successfully used in the two-meson systems,  but it was too costly for two-baryon systems until recently.  Such an approach  recently  led to the evidence for a weakly bound $H$ dibaryon in the SU(3) flavor limit and $m_\pi\simeq 420~$MeV \cite{Green:2021qol}.

Future variational studies including a wide range of interpolating operators, multiple lattice spacings, and the physical quark mass values will enable robust predictions of dibaryon binding energies, baryon-baryon  scattering amplitudes, and two-baryon matrix elements. Such LQCD calculations, and their generalizations to three-baryon and larger systems, will elucidate the quark and gluon structure of the lightest nuclei and predict poorly known baryon-baryon forces and two-body currents that appear as inputs to nuclear effective theories with a wide range of phenomenological applications.

     \subsection{Resonances with three-hadron decay channels}
\label{sec:threehadron}


Apart from coupled-channel scattering discussed in Sec.~\ref{sec:coupled}, another major challenge  for  hadron spectroscopy on the lattice is the determination of the properties of resonances with decay channels
involving three or more particles. 
For some resonances, such as the $a_1(1260)$, their three-body decay is dominant,
while for other states there are competing two- and three-body decay modes. 
Examples of the latter are almost all light baryonic resonances, including the Roper resonance discussed below,
for which both $\pi N$ and $\pi \pi N$ channels are available.
Other examples are
strange mesons that can decay to $\pi K$ and $\pi\pi K$~\cite{ParticleDataGroup:2020ssz}. 
Furthermore, while the light mesons discussed in previous sections usually couple to channels of two stable 
hadrons $(\pi\pi, \pi\eta, K\bar K,\dots)$, for heavier mesons the four-pion channel cannot be ignored. 
Understanding few-body dynamics from first principles is also crucial for 
multi-neutron forces relevant for the equation of state of neutron stars~\cite{Baym:2017whm}; 
recent advances in lattice QCD on few-nucleon systems~\cite{NPLQCD:2012mex, Savage:2016egr, Horz:2020zvv} complement dedicated experimental programs, such as at the FRIB facility~\cite{Gade:2016xrp}.

Extracting few-body scattering amplitudes from the lattice is challenging due to the many degrees of freedom present at a given total energy $E_{cm}$; this is in close analogy to coupled-channel scattering with spin discussed in Sec.~\ref{sec:coupled}. An example is given by the $\pi\rho$ system that can couple with different orbital angular momenta $l$ to the same $J^P$, even in the infinite volume~\cite{Woss:2018irj, Mai:2021nul}; indeed the $a_1(1260)$ can decay not only to these modes but also to $\pi f_0(500)$ and other channels~\cite{CLEO:1999rzk, ParticleDataGroup:2020ssz, Molina:2021awn}.

In light of this challenge,
a common strategy to determine infinite-volume amplitudes is to parametrize their energy dependence, which allows one to relate several measured eigenenergies from the lattice through a fit of a few parameters, and then evaluate the same amplitude in infinite volume. For the mapping from two- and three-body energy eigenvalues to physical, infinite-volume amplitudes, different \emph{quantization conditions} have been developed~\cite{Romero-Lopez:2021zdo}. For a comprehensive survey of the literature and a detailed comparison of different methods see recent reviews~\cite{Hansen:2019nir, Rusetsky:2019gyk, Mai:2021lwb}. 
Following the pioneering work of Ref.~\cite{Polejaeva:2012ut},
three major approaches have been followed, commonly denoted
the \emph{Relativistic Field Theory} (RFT)~\cite{Hansen:2014eka, Hansen:2015zga},  \emph{Non-Relativistic Effective Field Theory} (NREFT)~\cite{Hammer:2017uqm, Hammer:2017kms} and \emph{Finite Volume Unitarity} (FVU)~\cite{Mai:2017bge, Mai:2018djl} approaches. 
The NREFT approach was recently generalized to a manifestly relativistic-invariant form~\cite{Muller:2021uur}.
Similarities and equivalences exist between these approaches~\cite{Jackura:2019bmu, Briceno:2019muc, Blanton:2020gha, Blanton:2020jnm}.  
These three efforts are by no means the only ones in this rapidly expanding field, see, e.g., Refs.~\cite{Briceno:2012rv, Roca:2012rx, Konig:2017krd, Klos:2018sen, Romero-Lopez:2020rdq} as well as Refs.~\cite{Guo:2018ibd, Guo:2020kph} by Guo {\it et al.}, based on a variational approach and the Faddeev method.

Lattice QCD calculations of three-body systems have seen continuous progress in the last few years, pioneered by the NPLQCD collaboration, H\"orz/Hanlon, and the GWU-QCD, ETMC, and Hadron Spectrum collaborations for three-pion and three-kaon systems at maximal isospin~\cite{Beane:2007es, Detmold:2008fn, Detmold:2008yn, Blanton:2019vdk, Horz:2019rrn, Culver:2019vvu, Fischer:2020jzp, Hansen:2020otl, Alexandru:2020xqf}, and with the largest number of energy eigenvalues to date calculated in Ref.~\cite{Blanton:2021llb}. Usually, these calculations extend also to the two-body sectors forming subsystems of the three-body amplitude.

A major challenge is the extraction of the \emph{three-body force}. So far, it has been determined for the three-pion and three-kaon systems at maximal isospin including, in many cases, different quark masses~\cite{Detmold:2008fn, Mai:2018djl, Blanton:2019vdk, Culver:2019vvu, Blanton:2021llb, Fischer:2020jzp, Hansen:2020otl}; see, in particular, Ref.~\cite{Blanton:2021llb}, which was able to pin down also sub-dominant partial waves.  

This summary of current efforts leads naturally to the identification of  challenges for few-hadron physics from lattice QCD, which may be roughly ordered from short to long term as follows:
\begin{itemize}[leftmargin=*]
	\setlength{\itemsep}{0pt}
	\setlength{\parskip}{3pt}
	\setlength{\parsep}{0pt}
	\item 
        The three-body force, in general, is regularization-dependent.
Thus, the results, obtained in different approaches, should not be 
directly compared with each other or with the existing results of
theoretical calculations. Appropriate quantities to compare are the
$T$-matrix elements that are obtained through the solution of integral
equations in the infinite volume. In this connection, we also note
that the comparison of the latest results with the lowest-order chiral
perturbation theory predictions indicate a continued challenge to
understand even the weakly repulsive three-meson
systems at maximal isospin~\cite{Blanton:2021llb}.
	\item
	Strong three-body forces lead to resonance formation, which is especially relevant for the spectroscopy of the majority of meson and baryon resonances; a new lattice QCD calculation of the $a_1(1260)$ using three meson operators was recently carried out by the GWU-QCD collaboration~\cite{Mai:2021nul} (after earlier work in Ref.~\cite{Lang:2014tia} by Lang {\it et al.}). The extraction of energy eigenvalues was complemented by the mapping of the amplitude to the infinite volume using the FVU method. Analytic continuation of the three-body amplitude~\cite{Sadasivan:2021emk} allowed for the determination of the $a_1(1260)$ pole position and branching ratios (residues) to three-body channels with $\pi\rho$ quantum numbers in  partial waves $l\!=\!0$ and $2$. However, one needs more energy eigenvalues from the lattice to determine these resonance parameters more precisely and also to include more channels like $\pi f_0(500)$ that might contribute to the dynamics.
	\item
	This example points out the challenge that many three-body resonances decay to multiple partial waves/channels. The formalism for three-pion systems with different isospins in finite volume has been developed in the RFT formulation~\cite{Hansen:2020zhy, Blanton:2019igq} and provides a basis for future extensions to different quantum numbers. Both the RFT and FVU formalisms can now handle subsystems with spin and coupled channels. But the most general formulation for three-body systems in coupled channels, involving higher partial waves, moving frames, two- and three-body channels, and non-integer spin is still to be developed. 
	\item
	While using moving frames is standard for two-body systems, so as to obtain more energy eigenvalues, their inclusion increases admixtures of different values of $J^P$, thus complicating the analysis.
	For three particles, this problem is exacerbated as there can also be multiple three-particle channels present
	for a given $J^P$ of interest. 
Dealing with this will certainly require many eigenenergies in each frame, but may also require, as a first step,
calculations with pion masses 
sufficiently heavy that at least some channels can be neglected because of higher centrifugal barriers.
In addition, one might have to rely on phenomenological input from the PDG~\cite{ParticleDataGroup:2020ssz} or model calculations  (see, e.g., Refs.~\cite{Kamano:2011ih, MartinezTorres:2011vh,  Molina:2021awn}), at least until sufficiently many energy eigenvalues are available to drop some assumptions.
	\item
	Future applications of such slightly generalized finite-volume formalisms include the Roper resonance for which pioneering lattice QCD calculations have been performed albeit without three hadron operators~\cite{Lang:2016hnn}. Due to the presence of   $f_0(500) N$ in $\!l=\!0$ and other three-body channels~\cite{Ronchen:2012eg}, strong three-body effects for the finite-volume spectrum are expected.
	\item
	Predicting the properties of exotic mesons~\cite{Dudek:2010wm, Woss:2020ayi} is an important medium-term goal. Many of these decay substantially to three mesons~\cite{COMPASS:2009xrl, Meyer:2010ku}. But also ordinary mesons are of great interest such as the $\pi(1300)$ for which the three pions can all be in $l\!=\!0$~\cite{ParticleDataGroup:2020ssz} probably leading to strong three-body finite-volume effects.
	Such developments will provide theoretical support to ongoing experiments, e.g., at GlueX~\cite{GlueX:2020idb}, COMPASS~\cite{Ketzer:2019wmd}, and BESIII~\cite{BESIII:2020nme}.
	\item 
	Three-body decays represent the most direct access to three-body physics, an important example
	being $K\to3\pi$.  There is a pioneering calculation by the Hadron Spectrum Collaboration producing pseudo-Dalitz plots using lattice QCD~\cite{Hansen:2020otl}, but  finite-volume formalisms to more comprehensively study three-body decay amplitudes have only been developed recently, in the NREFT~\cite{Muller:2020wjo} and RFT approaches~\cite{Hansen:2021ofl}. Continuation along these lines is expected and relevant.
	\item 
	The quantization conditions provide a relationship between the finite-volume spectrum and unphysical quantities. The latter can be used to constrain the physical scattering amplitudes via a class of integral equation~\cite{Hansen:2015zga, Jackura:2018xnx}. To date, it is well understood how these equations may be solved for the lowest-lying partial waves in a restricted kinematic window~\cite{Hansen:2020otl, Jackura:2020bsk, Sadasivan:2021emk}. As the community continues to access increasingly complex reactions, it will be necessary to have a systematic procedure for solving these classes of integral equations for higher partial waves in the larger kinematic window, including in the complex plane of the different kinematic variables. Finally, these integral equations will aid in providing diagnostics of the different approaches for studying three-particle systems.
	\item
	Four-body dynamics is relevant for energies not too far above established resonances such as $\rho(770)$ and $f_0(500)$. Perturbative calculations for $n$-body systems in finite volume have been developed, even beyond the ground state~\cite{Beane:2007qr, Romero-Lopez:2020rdq}, but there is no general non-perturbative formalism available yet. Of course, the number of possible open channels is even larger than in the three-body case; a controllable starting point would be the system of four pions or kaons at maximal isospin.
\end{itemize}

 \subsection{High-lying resonances that will remain a challenge}\label{sec:unsolved}

Certain interesting resonances  will likely remain untouched challenges within current lattice simulation setups for some time. These are 
 the resonances that lie high above the lowest strong threshold and have many strong decay channels,  some of which contain 
 more than two hadrons. Listing just a few examples, these would be excited nucleons above $1.6~$GeV, higher-lying glueballs in dynamical QCD or the exotic $Z_c(4430)$ state.  It would be valuable  to explore alternative lattice approaches and to investigate whether certain simplifications can help with the extraction of the physical observables from lattice data.

 \section{Electro-weak transitions of resonances  }
 
The matrix elements $\langle H_f | J | H_i \rangle$  for   electroweak transitions between two strongly stable hadrons $H_i$ and $H_f$ have been explored on the lattice in great detail.  Many of them are crucial   to determine $V_{CKM}$ from the experimental data on the exclusive decays. The  lattice results on the corresponding form factors are reviewed for example in the FLAG review \cite{Aoki:2021kgd}.     
 Here we discuss the  electro-weak transitions of strongly decaying resonances, which represent the vast majority among hadrons. 
 
 The  transition $\langle R | J | H \rangle$ between the strongly stable hadron $H$ and a resonance $R$  via the electro-weak current  $J(q)$ needs to take the strong decay $R\to H_1H_2$ of a resonance into account.  The  transition  has to be investigated as a function of the momentum $q$  inserted by the current  and as a function of two-hadron energy $E_{H_1H_2}$.   The relation between   the matrix element extracted from a finite lattice and the one of interest for the infinite-volume continuum has been  analytically derived in
 \cite{Agadjanov:2014kha,Briceno:2014uqa,Briceno:2015csa,Agadjanov:2016fbd}, which builds on the previous ideas introduced in \cite{Lellouch:2000pv}. Only the electromagnetic transition $\langle \rho | J_{em} | \pi \rangle$ related to $\pi \gamma \to \rho\to  \pi
 \pi$ has been rigorously extracted from lattice QCD in this way \cite{Briceno:2015dca,Alexandrou:2018jbt}. It will be valuable to extract the electromagnetic transitions   $N \gamma \to \Delta \to  N \pi$ and $K \gamma \to K^* \to  K \pi$ in an analogous way. 
 The transitions between a stable hadron and a resonance via the weak external currents have never been rigorously extracted yet (with exception of $K\to \pi\pi$ discussed below). The first studies will likely consider the weak transitions  to the elastic resonance $\rho\to\pi\pi$, for example $D\to \rho l \bar \nu $
 and  $B\to \rho l \bar \nu $, which will serve also as test case to these challenging studies.   The lattice study of the weak decay $B\to K^* l^+l^- $ to an  elastic resonance $K^*\to K\pi$ and any  lepton pair $l=e,\mu,\tau$  will be of great  interest, since the experimental measurements hint to a possibility of new physics when comparing rates for various leptons in the final state.
  This lattice simulation is more challenging since it is induced by a number of currents and since $K^*$ is relatively narrow.   
   The decay $B\to D^* l\bar \nu$ to unstable $D^*\to D\pi$  should also be intensively investigated as the experimental rates indicate the violation of the lepton-flavor universality.  The $D^*$ is situated  very close to $D\pi$ threshold  in nature and it would be valuable to investigate whether this could influence the lattice results on the form factors.  
       
   The structure of resonances via the currents $\langle R | J | R \rangle$ can be rigorously explored using the formalism introduced in \cite{Bernard:2012bi,Briceno:2015tza, Baroni:2018iau, Briceno:2020vgp}, as discussed   Section \ref{sec:structure}.
   Alternatively, adapting the
      Feynman-Hellmann method of Ref.~\cite{QCDSF:2017ssq}
      to the calculation of the resonance matrix
      elements looks very promising.
 
  The only non-leptonic weak decay that has been explored in lattice QCD is   $K\to \pi\pi$.   The RBC/UKQCD made an impressive effort to study the kaon decays to $\pi\pi$  with isospins $I=0,2$ \cite{Blum:2015ywa,RBC:2020kdj} and also made the corresponding $\pi\pi$ scattering analysis \cite{RBC:2021acc} where the contribution of the $\sigma$ resonance is relevant in the isoscalar channel.

 \section{Towards the internal structure of hadrons}\label{sec:structure}
 
   Various interpretations for the existence of each exotic hadron have been proposed. Below we list certain observables, which might help to resolve their internal structure:
  \begin{enumerate}
 \item  Unlike experiment, lattice QCD can  explore how the properties of hadrons  depend on the quark masses of each flavor. 
 For this purpose, lattice simulations  should vary the quark masses and  determine the positions of the resonances and bound states with respect to their related thresholds.  If a hadron mass remains to be near the threshold   as the quark mass is varied, this could be interpreted as an indication of a sizable molecular   component.    A criterion for distinguishing the conventional and exotic states based on the quark mass dependence was proposed via the generalized Feynman-Hellmann theorem \cite{RuizdeElvira:2017aet}.  
  \item The structure of a stable hadron can be explored by evaluating the matrix elements $\langle H | J | H \rangle$. The currents $J $ thereby probe  the charge or energy densities of a given flavor in momentum or position space, for example.  In principle, also resonances $R$ can be explored via  $\langle R | J | R \rangle$ although this is significantly more challenging due to their decaying nature. The  formalism to extract these matrix elements from the lattice three-point functions was 
  introduced in \cite{Bernard:2012bi,Briceno:2015tza, Baroni:2018iau}. This formalism would allow for the determination of the scattering amplitude coupling two-hadron states via a current insertion for real-valued energies. The desired $\langle R | J | R \rangle$ matrix element can be obtained from the residue of this amplitude after it has been analytically continued to the resonance pole~\cite{Briceno:2020vgp}. This formalism has not been applied in lattice QCD simulations yet, and all the available  matrix elements of resonances have omitted their unstable nature  so far.    
  
   \item Recently diquark structure was explored by determining spatial quark density-density correlations $\langle O(t) \rho(\vec x_1,t')\rho(\vec x_2,t')O(0)\rangle$ for various current insertions $\rho=\bar q\Gamma q^\prime$ and for states   composed of a light diquark and a static heavy quark \cite{Francis:2021vrr}.  The dependence of these correlators on $\Gamma$, the distance $|\vec x_2 -\vec x_1|$ and the light quark flavor provide quantitative support  for an important role of the good diquark.  Application of a similar idea to other states would be of great interest. This would be relatively straightforward only for the strongly stable states.  
       
 \item  The importance of the molecular and diquark-antidiquark  Fock components within a  tetraquark with the minimal quark content $\bar q_1q_2 \bar q_3 q_4$ currently cannot be cleanly addressed since a rigorous criterion that would distinguish them is not available within the community.
  A clear distinction is further complicated since they are related via Fierz relations $\sum_n A_n (\bar q_1\Gamma_n^a q_2) (\bar q_3 \Gamma_n^b q_4)=\sum_m B_m [\bar q_1\Gamma_m^c \bar q_3]_{3_c} [q_2 \Gamma_m^d q_4]_{\bar 3_c}+C_m [\bar q_1\Gamma_m^e \bar q_3]_{6_c} [q_2 \Gamma_m^f q_4]_{\bar 6_c}$ which represent mathematical identities when  positions of quarks $q_i$  on both sides of the equations are the same.    This is a problem within any approach, not only lattice QCD, and highlights the need for further development.  
  
  \item  A careful analysis of lattice correlation functions can deliver the eigenenergies $E_n$ as well as the overlaps $\langle O_j|n\rangle$ of eigenstates $|n\rangle$ with the employed operators $O$ (\ref{C}). The qualitative information on the importance of various  Fock components in a given state  is often obtained from the overlaps  of an eigenstate $|n\rangle$ with the operators $O$ that resemble these Fock components.  
    However, it should be stressed that this does not lead to rigorous quantitative results on Fock components. This is because  the overlaps depend on a number of choices made by the lattice practitioner on the interpolating operators used. As such quark smearing as well as the renormalization scheme and scale obscure the connection. It would be valuable to explore whether some quantitative information on the nature of the states could be still extracted from the overlaps. Note that the overlaps of local operators can lead to rigorous information of the decay constants after the renormalization procedure.

 \end{enumerate}
 
 \section{Various   improvements}\label{sec:improvements}
 
Lattice  studies and the  related analytical considerations  would benefit from a variety of improvements, and some of them  are listed below:  
 \begin{enumerate}
 \item 
 Practically all the lattice QCD studies  of systems that contain pairs   $\bar cc$ or $\bar b b$ omit the    Wick contractions where $\bar QQ$ annihilates, with few exceptions, e.g. \cite{Knechtli:2021nth}.   This approximation is undertaken since many decay channels $\bar QQ\to H_1H_2...$ to   light hadrons  open up once the $\bar QQ$ annihilation is taken into account.  The effect of this annihilation on the   heavy quarkonia is expected to be   small due to the OZI-rule, e.g. \cite{Lepage:1992,Bodwin:1995}.   However, it would be valuable to devise ideas how to go beyond this approximation  in the future.  
 \item  Most of the scattering studies on the lattice are performed at a single value of the lattice spacing $a$ and the continuum limit is not taken.   Few simulations that extract the scattering information at several lattice spacings have been  done only recently. A study of $\Lambda\Lambda$ scattering in the H-dibaryon channel finds  non-negligible dependence of the 
energy-shifts  and  the scattering amplitude  on the lattice spacing \cite{Green:2021qol}. A significant effort in the future is needed  to explore the discretization   effects on various channels  that have been studied only at one lattice spacing.  
 \item 
 Practical improvements of the   methods to extract the poles of the scattering matrices from eigen-energies would be valuable. Here there are connections with techniques that have been developed to analyse experimental data.
 \item 
 Extraction of the  scattering amplitudes from the finite-volume spectral functions based on the LSZ formalism  could be attempted  \cite{Bulava:2019kbi}.
 \item One of the major problems in  investigating the higher-lying  states is that all eigenstates with the energy below that have to be extracted via the correlators (\ref{C}) for a certain quantum channel. It would be  valuable to find  an approach that could address just a certain higher-lying energy region. This does not seem viable with the currently used lattice methods.
 \end{enumerate}
 
  \section{ Conclusions} 
  Enormous progress has been made with ab-initio lattice QCD methods to study  masses and strong decay widths of various hadrons.  In spite of that, conclusions from this approach have not yet been drawn for many of the interesting exotic resonances  discovered in experiments. The main reason  is that most of these hadrons decay strongly via several decay channels. This makes the first-principle studies very challenging though not impossible.  It is expected that the lattice community will provide valuable results on certain exotic hadrons that are not situated too high above the lowest threshold.

 The impressive  discoveries on the experimental side will continue to motivate the progress on the theoretical side. 
   Resolving some of the listed    challenges would   improve our understanding of the experimentally observed conventional and exotic hadrons.  The predictions of yet-unobserved states will guide future experiments.   
   
   \vspace{1cm}

 {\bf Acknowledgments}
 
 \vspace{0.1cm}
 
M. D\"oring was    supported by the NSF grant PHY-2012289 and US. DOE  award DE-SC0016582.  
R. Lewis was supported in part by NSERC of Canada.
S. Prelovsek acknowledges support  by  ARRS  projects P1-0035 and J1-8137, and  DFG  research centre SFB/TRR-55.  
A. Rusetsky was supported by the  DFG  Project-ID 196253076 TRR 110,
Volkswagenstiftung  (grant 93562) and the Chinese Academy of Sciences President's International Fellowship Initiative (grant 2021VMB0007). 
S.R. Sharpe was supported in part by the U.S.  DOE grant   DE-SC0011637.
C.E. Thomas acknowledges support from the U.K. Science and Technology Facilities Council (STFC) grant ST/T000694/1.
M. Wagner acknowledges support by the  DFG  project  399217702.


\end{document}